\documentclass[prl,showpacs,twocolumn,amsmath,amssymb,superscriptaddress,groupedaddress,preprintnumbers]{revtex4}

\usepackage{amssymb}
\usepackage{graphicx}

\begin{document}

\title{Chaos and unpredictability in evolutionary dynamics in discrete time}

\author{Daniele Vilone}
\affiliation{Grupo Interdisciplinar de Sistemas Complejos (GISC), Departamento de Matem\'aticas, Universidad Carlos III de Madrid, 28911 Legan\'es, Spain}
\author{Alberto Robledo}

\affiliation{Grupo Interdisciplinar de Sistemas Complejos (GISC), Departamento de Matem\'aticas, Universidad Carlos III de Madrid, 28911 Legan\'es, Spain}

\affiliation{Instituto de F\'{\i}sica, Universidad Nacional Aut\'onoma 
de M\'exico, A. P. 20-364, 01000 M\'exico D. F., M\'exico}
\author{Angel S\'anchez}

\affiliation{Grupo Interdisciplinar de Sistemas Complejos (GISC), Departamento de Matem\'aticas, Universidad Carlos III de Madrid, 28911 Legan\'es, Spain}

\affiliation{Instituto de Biocomputaci\'on y F\'\i sica de Sistemas Complejos (BIFI), 
Universidad de Zaragoza, 50018 Zaragoza, Spain}

\begin{abstract}
A discrete-time version of the replicator equation for two-strategy games is studied. The stationary properties differ from those of continuous time for sufficiently large values of the parameters, where periodic and chaotic behavior replace the usual fixed-point population solutions. We observe the familiar period-doubling and chaotic-band-splitting attractor cascades of unimodal maps but in some cases more elaborate variations appear due to bimodality. Also unphysical stationary solutions can have unusual physical implications, such as the uncertainty of final population caused by sensitivity to initial conditions and fractality of attractor preimage manifolds.
\end{abstract}


\maketitle

Evolutionary dynamics deals with a generic situation in which individuals interact and 
reproduce according to the result of that interaction, which in turn depends on the 
composition of the population \cite{nowak:2006a,hofbauer:1998,hofbauer:2003}. 
This feedback loop gives rise to highly non trivial phenomena, 
that have attracted the interest of the physics community from the dynamical systems \cite{traulsen:2004,traulsen:2005,roca:2006,galla:2009} ,
statistical mechanics \cite{berg:1998,szabo:2002,floria:2009} and extended systems
 \cite{szabo:2007,castellano:2009,roca:2009a,perc:2010}
viewpoints. 
A particularly important class of problems is encompassed by 
the family generically referred to as {\em replicator dynamics} \cite{taylor:1978,maynard-smith:1982,schuster:1983}. If $x_i$ is the frequency of type $i=1,\ldots,n$, ${\bf x}\in\mathbb{R}^n$ 
is the frequency vector describing the population as a whole, and the 	
interaction is given by the $n\times n$ payoff matrix $A$, then $(A{\bf x})_i$ is the expected 
payoff for an individual of type $i$ and ${\bf x}^TA{\bf x}$ is the average payoff of the population. 
Assuming that the per capita rate of growth is given by the difference between the payoff
for type $i$ and the average
payoff in the population, we arrive at the replicator equation
\begin{equation}
\label{1}
\dot{x}^i=x_i[(A{\bf x})_i-{\bf x}^TA{\bf x}].
\end{equation}
Other forms have been proposed for this equation, which, e.g., include a denominator given
by the average global payoff (albeit in this case the orbits of the resulting equation are the
same) or other modifications (see, e.g., \cite{hofbauer:2003}; see also \cite{binmore:1995} and
references therein). However, Eq.\ (\ref{1}) is the 
one most often studied, and therefore we will stick to this form in the following. 

Among the very many nonlinear phenomena exhibited by the replicator equation, the 
appearance of chaotic 
dynamics has been studied in a number of papers. The Poincar\'e-Bendixson theorem \cite{schuster:1988} establishes that chaos can only arise in a continuous dynamical system (specified by differential equations) if it has three or more dimensions. Accordingly, Skyrms \cite{skyrms:1992} showed that 
chaos does not exist for three types (also referred to as strategies) and gave examples of
chaotic dynamics on strange attractors with four strategies. 
Further examples were provided in \cite{nowak:1993,chawanya:1995}.
Subsequently, 
Sato {\em et al.} \cite{sato:2002}
found both hamiltonian and non-hamiltonian chaos in a three-type 
version of the 
rock-scissors papers game \cite{nowak:2006a,maynard-smith:1982} 
by introducing asymmetry (payoffs depend on being the first or the second to interact).

In this letter we show that replicator dynamics does lead to chaos in games with only 
two types when the discretized version of the equation is considered, i.e., 
\begin{equation}
\label{2}
x^i_{t+1}=x^i_t+x^i_t[(A{\bf x_t})_i-({\bf x^t})^TA{\bf x_t}],
\end{equation}
where ${\bf x}_t$ is the composition of the population at time $t$. We will refer to Eq.\ 
(\ref{2}) in what follows as the {\em replicator mapping} and can be seen as arising from 
situations in which changes per generation are not necessarily small \cite{maynard-smith:1982}. 
We note that this is one of several
existing discrete versions of the replicator equations \cite{binmore:1995} and in particular
it has been considered in \cite{dekel:1992,cabrales:1992}. As we will see below, the relevance
of this result goes beyond the observation of chaos in low-dimensional systems under the required circumstances: Indeed,
the occurrence of chaos as well as of periodic trajectories gives rise to unexpected behavior
in well-known games for which so far only rest points have been reported. 

We will focus on the generic two-strategy game given by the payoff matrix
\begin{equation}
\begin{tabular}{|c|c|c|}
\hline
\mbox{ } & C & D 
 \\ \hline
C & 1 & S
 \\ \hline
D & T & 0 \\
\hline
\end{tabular}
\label{3}
\end{equation}
where payoffs for the row player are indicated. This choice
encompasses several of the most important dilemmas arising in both 
biological and social applications.
\begin{figure}
\includegraphics[width=8cm]{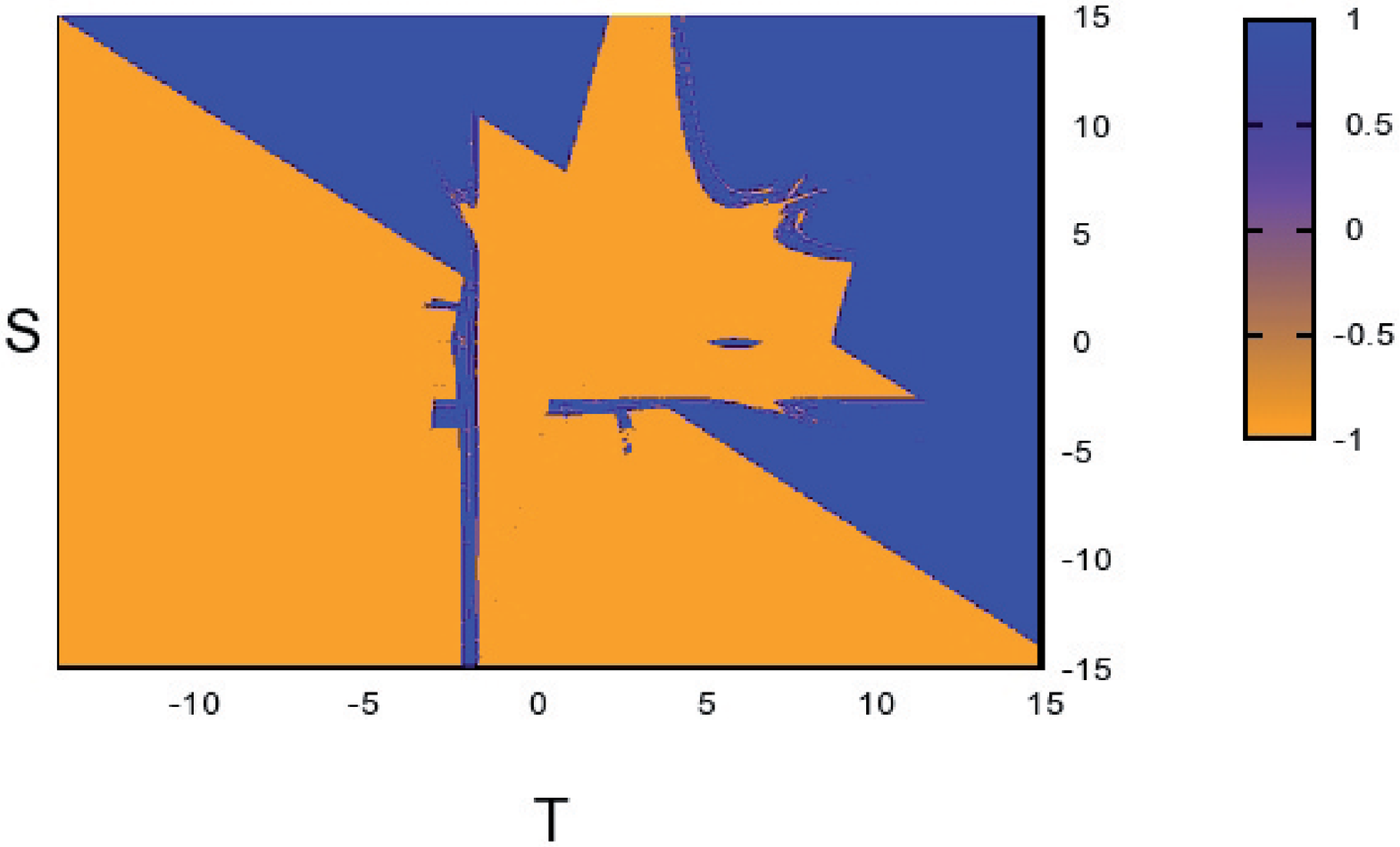}\\
\includegraphics[width=8cm]{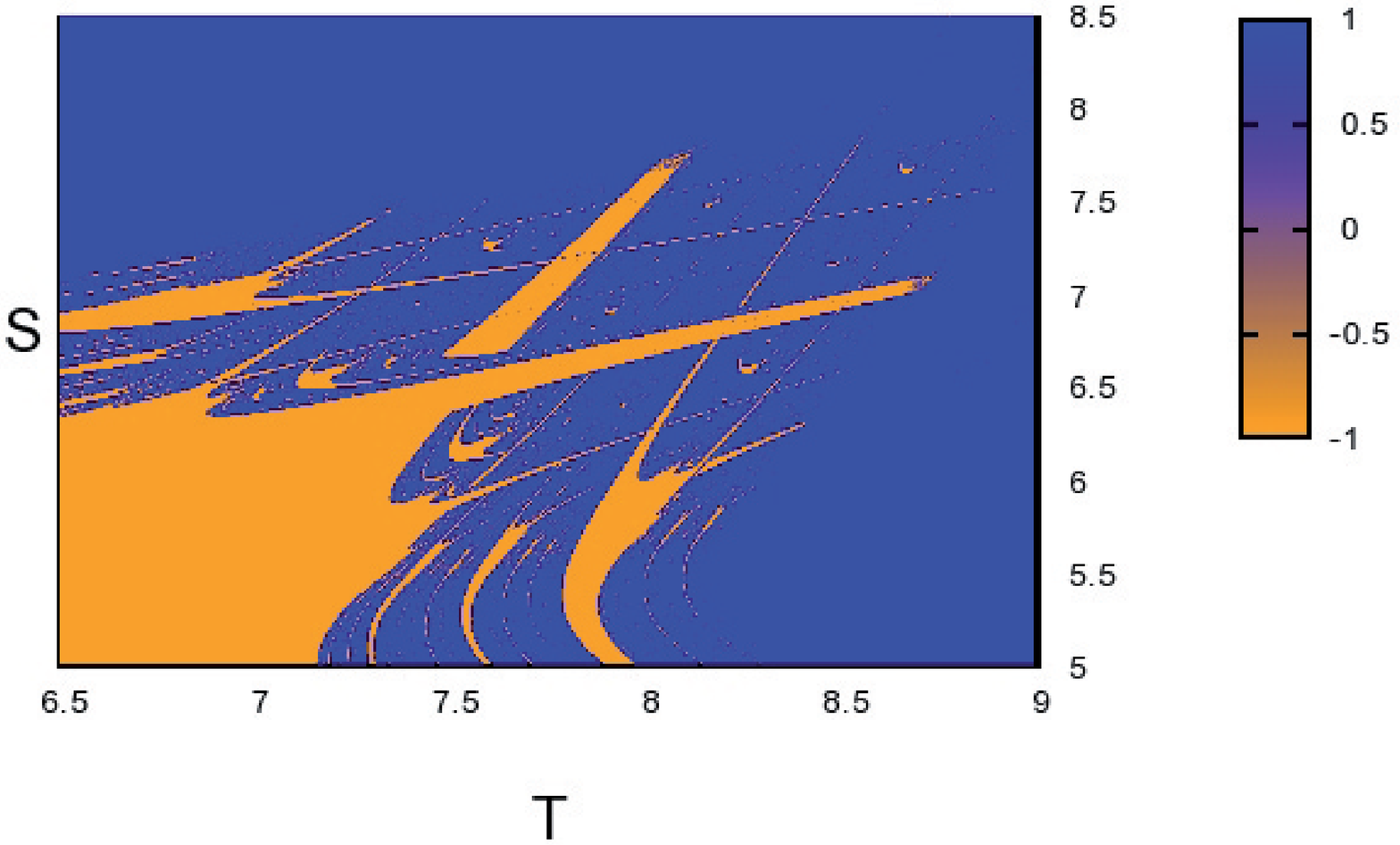}\\
\caption{
Upper panel: sign of the Lyapunov exponent (bright: negative, {\it i.e.} periodic solutions; dark: positive,
{\i.e.} chaotic solutions). Lower panel: zoom of a zone of the first region.}
\label{lyap_sign}
\end{figure}
According to the values
of $T$ and $S$, we have four basic social dilemmas: When $T>1$ and $S>0$ 
(upper-right region in Fig.~\ref{lyap_sign}) we obtain the Snowdrift game
(also known as Chicken or Hawk-Dove)
\cite{sugden:1986}, which represents a situation in which the best thing is to do the 
opposite of what the opponent does (i.e., it is an anti-coordination game). 
The region $T<1$ and $S<0$ (lower-left region)
corresponds to the Stag Hunt game \cite{skyrms:2003}, a coordination game in which what is best is to do
as the other player does. Finally, the other two regions define the Prisoner's 
Dilemma ($T>1$  and $S<0$, lower-right) \cite{axelrod:1984},
where D is the best option no matter what
the opponent chooses, and the Harmony game \cite{licht:1999}, 
with C being now the best strategy. In general, practically all the studies published on
this two-parametric representation of social dilemmas deal with the square $T\in[0,2],
S\in[-1,1]$ (referred to as ``basic square'' in what follows) 
as this is deemed enough to understand the four main classes of games,
although other special regions can be identified (see, e.g., \cite{hauert:2002}). 
With these payoffs, the replicator mapping given by Eq.\ (\ref{2}) becomes 
\begin{equation}
\label{4}
x_{t+1}=x_t+x_t(1-x_t)[S+(1-T-S)x_t],
\end{equation}
where $x_t$ now stands for the proportion of C-strategists in the population (the 
frequency of D-strategists being obviously $1-x_t$). 

To begin our analysis, we first consider the Lyapunov exponent of the mapping in Eq.\ (\ref{4}) as a function of 
$T$ and $S$, plotted in Figure~\ref{lyap_sign}. 
As can be seen from the figure, below the diagonal the Lyapunov exponent is negative
almost everywhere, whereas above the diagonal there is a region with intricate boundaries
where the exponent remains positive, and outside that region it becomes negative. 
Interestingly, the basic square is contained in the region in which the Lyapunov exponent
is negative. We recall that a negative Lyapunov exponent is indicative of 
periodic solutions and, as will be discussed below, periods $P>1$ do exist, although always 
outside the basic square, where only the well-known fixed-point solutions ($P=1$) are found. 
Perhaps this is the reason why all this new phenomenology has not been noticed before, and it
indicates that capturing the richness of evolutionary game theory, when viewed in discrete time,
demands the study of a larger payoff parameter region or, alternatively, of certain specific cases.
In fact, as the zoom in Fig.~\ref{lyap_sign}
shows, the boundary between the chaotic and non-chaotic regions is very complicated,
with fractal features, as it is regularly the case in dynamical systems. Another important remark is
that in relation to game theory we have observed two types of chaos. There is non-physical chaos when in the chaotic
region the variable $x_t$ becomes negative or takes values larger than 1, which are unphysical in so far it is
the frequency of C-strategists and therefore must be bounded by 0 and 1; this takes place
in the upper-left region, the Harmony game. On the other hand, in the
Snowdrift game, we have observed physical chaos in the sense that the values of $x_t$
remain bounded within 0 and 1 for all times, and therefore trajectories represent actually realizable evolutions
of populations. It is important to realize, however, that even non-physical chaos can be relevant in
the sense that while trajectories become unphysical above 1 or below 0, the existence of 
chaos makes unpredictable in practice where any initial condition will end, i.e., to which of
the two absorbing states $x_t=0$ or $x_t=1$ will become attached. Moreover, in case the chaotic motion takes place only around one of these two absorbing states,
so that we know exactly in which of them the system will freeze, the time $\tau$ needed
to reach the final configuration is sensitively dependent on the initial conditions and
therefore unpredictable. These considerations hold in particular when $\tau$ is large and
the system stays in the physical interval $[0,1]$ enough time before hitting a boundary: we have verified
that when the initial condition is close to an unstable equilibrium (or repellor), this is actually the case.

\begin{figure}
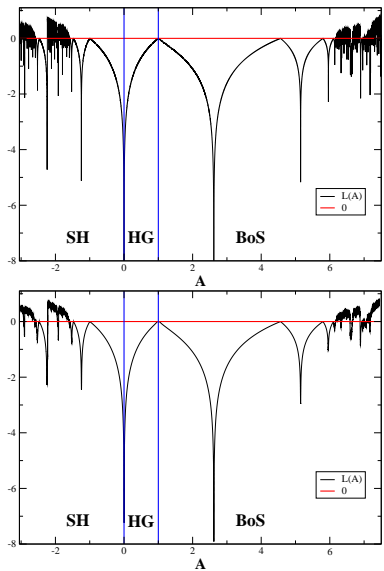

\includegraphics[clip=true,width=5cm]{HG+BoS_LyapE.eps}\\
\includegraphics[clip=true,width=5cm]{HG+BoS_LyapEnoise.eps}\\
\caption{
Upper panel: Behavior of the Lyapunov exponent along the diagonal $T=S$, as a function of the parameter $A=T=S$,
no noise. The acronyms indicate the games found in the different regions: SH, Stag Hunt; HG, 
Harmony Game; BoS, Battle of Sexes~\cite{traulsen:2005,skyrms:1992}. Lower panel: same graphics with noise ($\sigma=0.005$).}
\label{boslyap}
\end{figure}

\begin{figure}
\includegraphics[width=5.0cm]{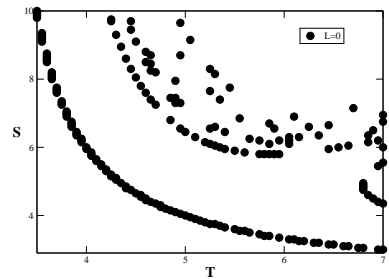}
\caption{Boundaries between areas of solutions with different periodicity. Proceeding away from the
bottom-left corner, dots indicate curves where the maximum Lyapunov exponent vanishes
and separate regions where the asymptotic state is a fixed point, a period-two
orbit, a period-four orbit, etc. The basic square lies outside the bottom-left corner of the plot,
in the fixed point region.}
\label{zero_lyap}
\end{figure}

\begin{figure}
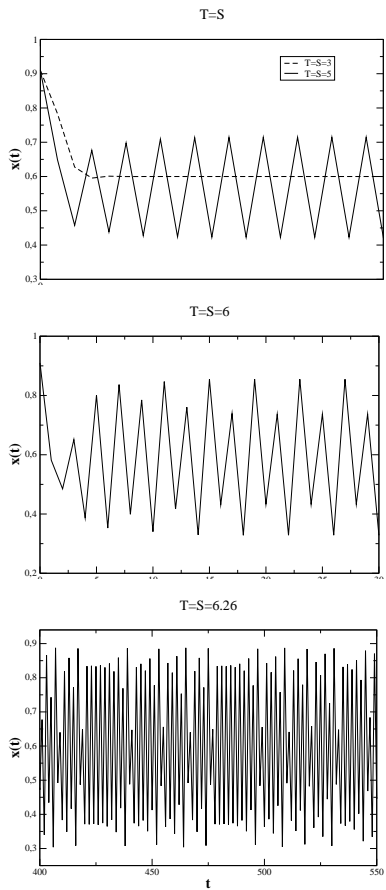

\includegraphics[width=5.0cm]{periodo1+2.eps}
\includegraphics[width=5.0cm]{periodo4.eps}
\includegraphics[width=5.0cm]{chaos.eps}
\caption{
Examples of time behavior for the map with $T=S$, with  $x_t$ being the cooperator density
vs time. From to top to bottom: periods one (fixed point) and two, period four, and chaotic
solution. Values of $T$ and $S$ indicated.}
\label{dyn_ex}
\end{figure}

In order to gain more insight on how the transition to chaos takes place, we have focused 
on the case $T=S=A$ and represented the Lyapunov exponent  as a function
of the new parameter $A$. Figure~\ref{boslyap} shows the behavior of the 
Lyapunov exponent as a function of $A$. The upper plot clearly shows that both in the
region where we found physical ($A>0$) or
unphysical ($A<0$) behavior there is a well defined route to chaos. Furthermore, the unphysical
region reproduces the period-doubling route to chaos characteristic of unimodal (one hump) maps, whereas 
in the physical region we have a different but not unrelated route that arises because of bimodality (two extrema) of the map when $A>0$. 
For small to moderate $A>0$ trajectories settle into periodic and chaotic attractors
generated separately by each of the extrema of the map, but for sufficiently large
$A$ trajectories spread and bounce between both extrema and converge to new periodic
and chaotic attractors. This is the reason why in Fig.~\ref{boslyap} the Lyapunov
exponent for large $A>0$ differs from the familiar unimodal pattern shown when $A<0$.
The lower panel of Fig.~\ref{boslyap} shows the effect on the Lyapunov exponent of additive
noise in Eq.~(\ref{4}). As expected, we observe a gradual smearing out of the fine structure
due to the removal of periodic and chaotic-band attractors with increasing noise amplitude
$\sigma$. This amplitude fixes the largest period and number of bands present, $2^N(\sigma)$,
giving rise to the so-called bifurcation gap~\cite{crutchfield:1982}.
Importantly, the chaotic behavior is not restricted to the diagonal: As shown in
Fig.~\ref{zero_lyap} the boundary between the fixed point solution
(the same as in the continuous time replicator equation) and the period-two orbits takes a
hyperbole-like shape. Then, moving outwards in the $T-S$ plane, boundaries between
period two and period four orbits appear, and so on. It is important to note that the plane
region usually considered in these studies, namely the basic square
$T\in [0,2],\ \ S\in [-1,1]$ (bottom left of, and out of, Fig.~\ref{zero_lyap}) contains only fixed point
solutions, i.e., in that region the continuous replicator equation and the discretized version
we are studying here lead to the same dynamical behavior. 
Finally, Fig.~\ref{dyn_ex} illustrates the dynamics typically found in the route to chaos along
the diagonal ($T=S$) of the region with physical behavior. As $A$ is increased, we observe 
the convergence to the fixed point, continuous-like solution,
or to periodic solutions until reaching chaos (bottom panel of Fig.~\ref{dyn_ex}). This route to chaos appears an infinite number of times along the family of attractors generated by unimodal maps within periodic windows that interrupt sections of chaotic attractors. In the opposite parameter direction, a route out of chaos accompanies each period-doubling cascade by a chaotic band splitting cascade \cite{schuster:1988}. Bimodal maps with a minimum and a maximum (as in the case $A>0$) display interesting variations of these attractor cascades. 

In summary, we have studied a discretized version of the replicator dynamics and found that 
there are large regions of the $T-S$ parameter space of the typical social dilemmas for which
considering discrete time leads to behavior largely different from the continuous one. 
Indeed, beyond the restricted region which has been analyzed before, we observe that
as the parameters $T$ and $S$ increase, periodic ($P>1$) and chaotic dynamics occurs.
This is an important difference with respect to 
the continuous time dynamics, for which only fixed point solutions are found.  We have 
also classified those novel behaviors as physical and unphysical, depending on whether
the values of the concentration of C-strategists remain in $[0,1]$ or spread beyond the 
boundaries of this interval. In the physical case, periodic solutions imply that C and 
D strategists vary periodically in frequency, the state of the system being generally 
mixed, containing both types of individuals. Such cycles have only been observed
in the continuous replicator dynamics for rock-scissors-paper games, i.e., games with
three strategies \cite{nowak:2006a,hofbauer:1998,hofbauer:2003}. The occurrence
of chaotic behavior adds unpredictability to the features of the system, in complete contrast
with the predictive character of the continuous case. On the other hand, periodic or chaotic 
behavior, even being unphysical, still has important consequences: In the time 
interval in which the trajectory is still contained in $[0,1]$, in the chaotic region it turns out
that it is not possible to predict whether, starting from a given initial condition, the 
system will end up in a population fully consisting of C- or D-strategists. We stress that
our results may be relevant for real life applications, in many cases involving 
bacteria, animals or humans which are well described by social dilemmas that do not have
well-determined payoffs, and often take place in discrete time. In principle, it is possible
to test our ideas in experimental setups, at least with human volunteers, by setting 
properly chosen payoffs and interactions as discrete events. Clearly, the appearance
of periodic orbits and chaos is generic and will affect other 
strategic interactions when described by a discretized equation. Unimodal maps (obtained e.g., when $A<0$) show a rich self-affine structure of chaotic-band attractors interspersed by periodic windows within which multiple period-doubling and band-splitting cascades take place. Bimodal maps (obtained e.g., when $A>0$) display more elaborate attractor structures. These dynamical features manifest in the discrete-time evolutionary dynamics addressed here and translate into matching patterns of population behavior.

\enlargethispage{4mm}

 
 D.\ V.\ aknowledges a postdoctoral contract from Universidad Carlos III de
 Madrid. 
  The stay of A.\ R.\ at
Universidad Carlos III de Madrid was supported by a grant from the Ministerio de 
Educaci\'on y Ciencia (Spain). 
A.\ S.\ aknowledges grants MOSAICO and Complexity-NET RESINEE 
(Ministerio de Ciencia e Innovaci\'on,
Spain) and MODELICO-CM (Comunidad de Madrid, Spain).

\bibliography{biblio}

\begin{thebibliography}{31}
\expandafter\ifx\csname natexlab\endcsname\relax\def\natexlab#1{#1}\fi
\expandafter\ifx\csname bibnamefont\endcsname\relax
  \def\bibnamefont#1{#1}\fi
\expandafter\ifx\csname bibfnamefont\endcsname\relax
  \def\bibfnamefont#1{#1}\fi
\expandafter\ifx\csname citenamefont\endcsname\relax
  \def\citenamefont#1{#1}\fi
\expandafter\ifx\csname url\endcsname\relax
  \def\url#1{\texttt{#1}}\fi
\expandafter\ifx\csname urlprefix\endcsname\relax\def\urlprefix{URL }\fi
\providecommand{\bibinfo}[2]{#2}
\providecommand{\eprint}[2][]{\url{#2}}

\bibitem[{\citenamefont{Nowak}(2006)}]{nowak:2006a}
\bibinfo{author}{\bibfnamefont{M.~A.} \bibnamefont{Nowak}},
  \emph{\bibinfo{title}{Evolutionary dynamics: exploring the equations of
  life}} (\bibinfo{publisher}{The Belknap Press of Harvard University Press},
  \bibinfo{address}{Cambridge}, \bibinfo{year}{2006}).

\bibitem[{\citenamefont{Hofbauer and Sigmund}(1998)}]{hofbauer:1998}
\bibinfo{author}{\bibfnamefont{J.}~\bibnamefont{Hofbauer}} \bibnamefont{and}
  \bibinfo{author}{\bibfnamefont{K.}~\bibnamefont{Sigmund}},
  \emph{\bibinfo{title}{Evolutionary Games and Population Dynamics}}
  (\bibinfo{publisher}{Cambridge University Press},
  \bibinfo{address}{Cambridge}, \bibinfo{year}{1998}).

\bibitem[{\citenamefont{Hofbauer and Sigmund}(2003)}]{hofbauer:2003}
\bibinfo{author}{\bibfnamefont{J.}~\bibnamefont{Hofbauer}} \bibnamefont{and}
  \bibinfo{author}{\bibfnamefont{K.}~\bibnamefont{Sigmund}},
  \bibinfo{journal}{Bulletin of the American Mathematical Society}
  \textbf{\bibinfo{volume}{40}}, \bibinfo{pages}{479} (\bibinfo{year}{2003}).

\bibitem[{\citenamefont{Traulsen et~al.}(2004)\citenamefont{Traulsen, R{\"o}hl,
  and Schuster}}]{traulsen:2004}
\bibinfo{author}{\bibfnamefont{A.}~\bibnamefont{Traulsen}},
  \bibinfo{author}{\bibfnamefont{T.}~\bibnamefont{R{\"o}hl}}, \bibnamefont{and}
  \bibinfo{author}{\bibfnamefont{H.~G.} \bibnamefont{Schuster}},
  \bibinfo{journal}{Phys. Rev. Lett.} \textbf{\bibinfo{volume}{93}},
  \bibinfo{pages}{28701} (\bibinfo{year}{2004}).

\bibitem[{\citenamefont{Traulsen et~al.}(2005)\citenamefont{Traulsen, Claussen,
  and Hauert}}]{traulsen:2005}
\bibinfo{author}{\bibfnamefont{A.}~\bibnamefont{Traulsen}},
  \bibinfo{author}{\bibfnamefont{J.~C.} \bibnamefont{Claussen}},
  \bibnamefont{and} \bibinfo{author}{\bibfnamefont{C.}~\bibnamefont{Hauert}},
  \bibinfo{journal}{Phys. Rev. Lett.} \textbf{\bibinfo{volume}{95}},
  \bibinfo{pages}{238701} (\bibinfo{year}{2005}).

\bibitem[{\citenamefont{Roca et~al.}(2006)\citenamefont{Roca, Cuesta, and
  S\'anchez}}]{roca:2006}
\bibinfo{author}{\bibfnamefont{C.~P.} \bibnamefont{Roca}},
  \bibinfo{author}{\bibfnamefont{J.~A.} \bibnamefont{Cuesta}},
  \bibnamefont{and}
  \bibinfo{author}{\bibfnamefont{A.}~\bibnamefont{S\'anchez}},
  \bibinfo{journal}{Phys. Rev. Lett.} \textbf{\bibinfo{volume}{97}},
  \bibinfo{pages}{158701} (\bibinfo{year}{2006}).

\bibitem[{\citenamefont{Galla}(2009)}]{galla:2009}
\bibinfo{author}{\bibfnamefont{T.}~\bibnamefont{Galla}},
  \bibinfo{journal}{Phys. Rev. Lett.} \textbf{\bibinfo{volume}{103}},
  \bibinfo{pages}{198702} (\bibinfo{year}{2009}).

\bibitem[{\citenamefont{Berg and Engel}(1998)}]{berg:1998}
\bibinfo{author}{\bibfnamefont{J.}~\bibnamefont{Berg}} \bibnamefont{and}
  \bibinfo{author}{\bibfnamefont{A.}~\bibnamefont{Engel}},
  \bibinfo{journal}{Phys. Rev. Lett.} \textbf{\bibinfo{volume}{81}},
  \bibinfo{pages}{4999} (\bibinfo{year}{1998}).

\bibitem[{\citenamefont{Szab{\'o} and Hauert}(2002)}]{szabo:2002}
\bibinfo{author}{\bibfnamefont{G.}~\bibnamefont{Szab{\'o}}} \bibnamefont{and}
  \bibinfo{author}{\bibfnamefont{C.}~\bibnamefont{Hauert}},
  \bibinfo{journal}{Phys. Rev. Lett.} \textbf{\bibinfo{volume}{89}},
  \bibinfo{pages}{118101} (\bibinfo{year}{2002}).

\bibitem[{\citenamefont{Flor{\'\i}a et~al.}(2009)\citenamefont{Flor{\'\i}a,
  Gracia-Lazaro, Garde{\~n}es, and Moreno}}]{floria:2009}
\bibinfo{author}{\bibfnamefont{L.~M.} \bibnamefont{Flor{\'\i}a}},
  \bibinfo{author}{\bibfnamefont{C.}~\bibnamefont{Gracia-Lazaro}},
  \bibinfo{author}{\bibfnamefont{J.~G.} \bibnamefont{Garde{\~n}es}},
  \bibnamefont{and} \bibinfo{author}{\bibfnamefont{Y.}~\bibnamefont{Moreno}},
  \bibinfo{journal}{Phys. Rev. E} \textbf{\bibinfo{volume}{79}},
  \bibinfo{pages}{026106} (\bibinfo{year}{2009}).

\bibitem[{\citenamefont{Szab{\'o} and F{\'a}th}(2007)}]{szabo:2007}
\bibinfo{author}{\bibfnamefont{G.}~\bibnamefont{Szab{\'o}}} \bibnamefont{and}
  \bibinfo{author}{\bibfnamefont{G.}~\bibnamefont{F{\'a}th}},
  \bibinfo{journal}{Phys.\ Rep.} \textbf{\bibinfo{volume}{446}},
  \bibinfo{pages}{97} (\bibinfo{year}{2007}).

\bibitem[{\citenamefont{Castellano et~al.}(2009)\citenamefont{Castellano,
  Fortunato, and Loreto}}]{castellano:2009}
\bibinfo{author}{\bibfnamefont{C.}~\bibnamefont{Castellano}},
  \bibinfo{author}{\bibfnamefont{S.}~\bibnamefont{Fortunato}},
  \bibnamefont{and} \bibinfo{author}{\bibfnamefont{V.}~\bibnamefont{Loreto}},
  \bibinfo{journal}{Rev. Mod. Phys.} \textbf{\bibinfo{volume}{81}},
  \bibinfo{pages}{591} (\bibinfo{year}{2009}).

\bibitem[{\citenamefont{Roca et~al.}(2009)\citenamefont{Roca, Cuesta, and
  S{\'a}nchez}}]{roca:2009a}
\bibinfo{author}{\bibfnamefont{C.~P.} \bibnamefont{Roca}},
  \bibinfo{author}{\bibfnamefont{J.}~\bibnamefont{Cuesta}}, \bibnamefont{and}
  \bibinfo{author}{\bibfnamefont{A.}~\bibnamefont{S{\'a}nchez}},
  \bibinfo{journal}{Phys. Life Rev.} \textbf{\bibinfo{volume}{6}},
  \bibinfo{pages}{208} (\bibinfo{year}{2009}).

\bibitem[{\citenamefont{Perc and Szolnoki}(2010)}]{perc:2010}
\bibinfo{author}{\bibfnamefont{M.}~\bibnamefont{Perc}} \bibnamefont{and}
  \bibinfo{author}{\bibfnamefont{A.}~\bibnamefont{Szolnoki}},
  \bibinfo{journal}{BioSystems} \textbf{\bibinfo{volume}{99}},
  \bibinfo{pages}{109} (\bibinfo{year}{2010}).

\bibitem[{\citenamefont{Taylor and Jonker}(1978)}]{taylor:1978}
\bibinfo{author}{\bibfnamefont{P.~D.} \bibnamefont{Taylor}} \bibnamefont{and}
  \bibinfo{author}{\bibfnamefont{L.}~\bibnamefont{Jonker}},
  \bibinfo{journal}{Math.\ Biosci.} \textbf{\bibinfo{volume}{40}},
  \bibinfo{pages}{145} (\bibinfo{year}{1978}).

\bibitem[{\citenamefont{{Maynard Smith}}(1982)}]{maynard-smith:1982}
\bibinfo{author}{\bibfnamefont{J.}~\bibnamefont{{Maynard Smith}}},
  \emph{\bibinfo{title}{Evolution and the Theory of Games}}
  (\bibinfo{publisher}{Cambridge University Press},
  \bibinfo{address}{Cambridge}, \bibinfo{year}{1982}).

\bibitem[{\citenamefont{Schuster and Sigmund}(1983)}]{schuster:1983}
\bibinfo{author}{\bibfnamefont{P.}~\bibnamefont{Schuster}} \bibnamefont{and}
  \bibinfo{author}{\bibfnamefont{K.}~\bibnamefont{Sigmund}},
  \bibinfo{journal}{J. Theor. Biol.} \textbf{\bibinfo{volume}{100}},
  \bibinfo{pages}{533} (\bibinfo{year}{1983}).

\bibitem[{\citenamefont{Binmore et~al.}(1995)\citenamefont{Binmore, Samuelson,
  and Vaughan}}]{binmore:1995}
\bibinfo{author}{\bibfnamefont{K.~G.} \bibnamefont{Binmore}},
  \bibinfo{author}{\bibfnamefont{L.}~\bibnamefont{Samuelson}},
  \bibnamefont{and} \bibinfo{author}{\bibfnamefont{R.}~\bibnamefont{Vaughan}},
  \bibinfo{journal}{Games Econ. Behav.} \textbf{\bibinfo{volume}{11}},
  \bibinfo{pages}{1} (\bibinfo{year}{1995}).

\bibitem[{\citenamefont{Schuster}(1988)}]{schuster:1988}
\bibinfo{author}{\bibfnamefont{H.}~\bibnamefont{Schuster}},
  \emph{\bibinfo{title}{Deterministic Chaos. An Introduction, 2nd revised ed.}}
  (\bibinfo{publisher}{VCH, Weinheim}, \bibinfo{year}{1988}).

\bibitem[{\citenamefont{Skyrms}(1992)}]{skyrms:1992}
\bibinfo{author}{\bibfnamefont{B.}~\bibnamefont{Skyrms}}, \bibinfo{journal}{J.
  Logic Lang. Infor.} \textbf{\bibinfo{volume}{1}}, \bibinfo{pages}{111}
  (\bibinfo{year}{1992}).

\bibitem[{\citenamefont{Nowak and Sigmund}(1993)}]{nowak:1993}
\bibinfo{author}{\bibfnamefont{M.~A.} \bibnamefont{Nowak}} \bibnamefont{and}
  \bibinfo{author}{\bibfnamefont{K.}~\bibnamefont{Sigmund}},
  \bibinfo{journal}{Proc. Natl. Acad. Sci. USA} \textbf{\bibinfo{volume}{90}},
  \bibinfo{pages}{5091} (\bibinfo{year}{1993}).

\bibitem[{\citenamefont{Chawanya}(1995)}]{chawanya:1995}
\bibinfo{author}{\bibfnamefont{T.}~\bibnamefont{Chawanya}},
  \bibinfo{journal}{Prog. Theor. Phys.} \textbf{\bibinfo{volume}{94}},
  \bibinfo{pages}{163} (\bibinfo{year}{1995}).

\bibitem[{\citenamefont{Sato et~al.}(2002)\citenamefont{Sato, Akiyama, and
  Farmer}}]{sato:2002}
\bibinfo{author}{\bibfnamefont{Y.}~\bibnamefont{Sato}},
  \bibinfo{author}{\bibfnamefont{E.}~\bibnamefont{Akiyama}}, \bibnamefont{and}
  \bibinfo{author}{\bibfnamefont{J.~D.} \bibnamefont{Farmer}},
  \bibinfo{journal}{Proc. Natl. Acad. Sci. USA} \textbf{\bibinfo{volume}{99}},
  \bibinfo{pages}{4748} (\bibinfo{year}{2002}).

\bibitem[{\citenamefont{Dekel and Scotchmer}(1992)}]{dekel:1992}
\bibinfo{author}{\bibfnamefont{E.}~\bibnamefont{Dekel}} \bibnamefont{and}
  \bibinfo{author}{\bibfnamefont{S.}~\bibnamefont{Scotchmer}},
  \bibinfo{journal}{J.\ Econ.\ Theory} \textbf{\bibinfo{volume}{57}},
  \bibinfo{pages}{392} (\bibinfo{year}{1992}).

\bibitem[{\citenamefont{Cabrales and Sobel}(1992)}]{cabrales:1992}
\bibinfo{author}{\bibfnamefont{A.}~\bibnamefont{Cabrales}} \bibnamefont{and}
  \bibinfo{author}{\bibfnamefont{J.}~\bibnamefont{Sobel}},
  \bibinfo{journal}{J.\ Econ.\ Theory} \textbf{\bibinfo{volume}{57}},
  \bibinfo{pages}{407} (\bibinfo{year}{1992}).

\bibitem[{\citenamefont{Sugden}(1986)}]{sugden:1986}
\bibinfo{author}{\bibfnamefont{R.}~\bibnamefont{Sugden}},
  \emph{\bibinfo{title}{The economics of rights, co-operation and welfare}}
  (\bibinfo{publisher}{Basil Blackwell}, \bibinfo{address}{Oxford},
  \bibinfo{year}{1986}).

\bibitem[{\citenamefont{Skyrms}(2003)}]{skyrms:2003}
\bibinfo{author}{\bibfnamefont{B.}~\bibnamefont{Skyrms}},
  \emph{\bibinfo{title}{The Stag Hunt and Evolution of Social Structure}}
  (\bibinfo{publisher}{Cambridge University Press},
  \bibinfo{address}{Cambridge}, \bibinfo{year}{2003}).

\bibitem[{\citenamefont{Axelrod}(1984)}]{axelrod:1984}
\bibinfo{author}{\bibfnamefont{R.}~\bibnamefont{Axelrod}},
  \emph{\bibinfo{title}{The Evolution of Cooperation}}
  (\bibinfo{publisher}{Basic Books}, \bibinfo{address}{New York},
  \bibinfo{year}{1984}).

\bibitem[{\citenamefont{Licht}(1999)}]{licht:1999}
\bibinfo{author}{\bibfnamefont{A.~N.} \bibnamefont{Licht}},
  \bibinfo{journal}{Yale J.\ Int.\ Law} \textbf{\bibinfo{volume}{24}},
  \bibinfo{pages}{61} (\bibinfo{year}{1999}).

\bibitem[{\citenamefont{Hauert}(2002)}]{hauert:2002}
\bibinfo{author}{\bibfnamefont{C.}~\bibnamefont{Hauert}},
  \bibinfo{journal}{Int.\ J.\ Bif.\ Chaos} \textbf{\bibinfo{volume}{12}},
  \bibinfo{pages}{1531} (\bibinfo{year}{2002}).

\bibitem[{\citenamefont{Crutchfield et~al.}(1982)\citenamefont{Crutchfield,
  Farmer, and Huberman}}]{crutchfield:1982}
\bibinfo{author}{\bibfnamefont{J.~P.} \bibnamefont{Crutchfield}},
  \bibinfo{author}{\bibfnamefont{J.}~\bibnamefont{Farmer}}, \bibnamefont{and}
  \bibinfo{author}{\bibfnamefont{B.~A.} \bibnamefont{Huberman}},
  \bibinfo{journal}{Physics Reports} \textbf{\bibinfo{volume}{92}},
  \bibinfo{pages}{45} (\bibinfo{year}{1982}).

\end{thebibliography}

\end{document}